\documentclass[conference]{IEEEtran}
\usepackage[T1]{fontenc}
\usepackage{graphicx}
\usepackage{cite}
\usepackage{subcaption}

\usepackage{overpic}
\usepackage{amssymb}
\usepackage{amsmath}
\usepackage{amsthm}
\usepackage{steinmetz}
\usepackage{array}
\usepackage{lipsum}  
\usepackage{url}
\usepackage[dvipsnames]{xcolor}

\theoremstyle{plain}

\newtheorem{lemma}{Lemma}

\newtheorem{corollary}{Corollary}

\newtheorem{proposition}{Proposition}
\newtheorem{assumption}{Assumption}

\newcommand{\vect}[1]{\mathbf{#1}}

\def\Htran{\mbox{\tiny $\mathrm{H}$}}
\def\Ttran{\mbox{\tiny $\mathrm{T}$}}
\def\CN{\mathcal{N}_{\mathbb{C}}} 

\def\mod{\mathrm{mod}}

\begin{document}

\title{Near- and Far-Field Communications with \\ Large Intelligent Surfaces \vspace{-0.4cm}}

\author{
\IEEEauthorblockN{Andrea de Jesus Torres\IEEEauthorrefmark{1}, Luca Sanguinetti\IEEEauthorrefmark{1},
Emil Bj{\"o}rnson\IEEEauthorrefmark{2}
\thanks{L. Sanguinetti was partially supported by the Italian Ministry of Education and Research (MIUR) in the framework of the CrossLab project (Departments of Excellence).}}
\IEEEauthorblockA{\IEEEauthorrefmark{1}\small{Dipartimento di Ingegneria dell'Informazione, University of Pisa, Italy (andrea.dejesustorres@phd.unipi.it, luca.sanguinetti@unipi.it)}}
\IEEEauthorblockA{\IEEEauthorrefmark{2}\small{KTH Royal Institute of Technology and Link\"oping University, Sweden (emilbjo@kth.se) }\vspace{-0.3cm}}
}

%
%

\maketitle
\begin{abstract}
This paper studies the uplink spectral efficiency (SE) achieved by two single-antenna user equipments (UEs) communicating with a Large Intelligent Surface (LIS), defined as a planar array consisting of $N$ antennas that each has area $A$. The analysis is carried out with a deterministic line-of-sight propagation channel model that captures key fundamental aspects of the so-called geometric near-field of the array. Maximum ratio (MR) and minimum mean squared error (MMSE) combining schemes are considered. With both schemes, the signal and interference terms are numerically analyzed as a function of the position of the transmitting devices when the width/height $L = \sqrt{NA}$ of the square-shaped array grows large. The results show that an exact near-field channel model is needed to evaluate the SE whenever the distance of transmitting UEs is comparable with the LIS' dimensions. It is shown that, if $L$ grows, the UEs are eventually in the geometric near-field and the interference does not vanish. MMSE outperforms MR for an LIS of practically large size.
\end{abstract}
\smallskip
\section{Introduction}
Large Intelligent Surface (LIS) refers to arrays with a massive number of antennas in a compact space~\cite{Hu2018a}. In its asymptotic form, it can be thought of as a spatially-continuous electromagnetic aperture that actively generates beamformed radio signals or receives them accordingly. Research on this topic is performed under many different names~\cite{zhao2019survey}, among them: holographic MIMO~\cite{Pizzo2020}; reconfigurable intelligent surface~\cite{Garcia2019a}; and software-defined surface (SDS)~\cite{basar2020reconfigurable}. A common practice in multiple antenna communications is to approximate the received electromagnetic wave with a plane wave. This approximation is valid when the terminal distance is much larger than the dimensions of the array and brings to the well-known geometric far-field approximation. This paper investigates the potential deficiencies of this approximation when an LIS of size comparable to (or larger than) the distance from the transmitting devices is considered. For brevity, we consider the deterministic line-of-sight propagation channel from~\cite{bjornson2020power,Dardari2019a}, which allows to study the so-called geometric near-field of the array by taking into account three fundamental aspects: 1) the varying distance to the antennas in the LIS; 2) the varying effective antenna areas; 3) the varying loss from polarization mismatch. Unlike~\cite{bjornson2020power,Dardari2019a}, this channel model is used to evaluate the uplink spectral efficiency (SE) achieved by two single-antenna user equipments (UEs) when communicating with an LIS of square geometry, using either maximum ratio (MR) or minimum mean squared error (MMSE) combining. Comparisons with the far-field approximation will show that the exact near-field model is unarguably needed when the LIS has comparable size to UE distances. We will also show that, as the LIS size grows, the UEs will eventually be in the geometric near-field and the interference will not vanish, which is different from what is conventionally considered in the Massive MIMO literature. Moreover, it will be shown that MMSE provides performance that is superior to MR for an LIS of practically large size. This makes it the preferred combining scheme.

\begin{figure}[t!]\hspace{0.25cm}
        \centering 
	\begin{overpic}[width=0.6\columnwidth,tics=10]{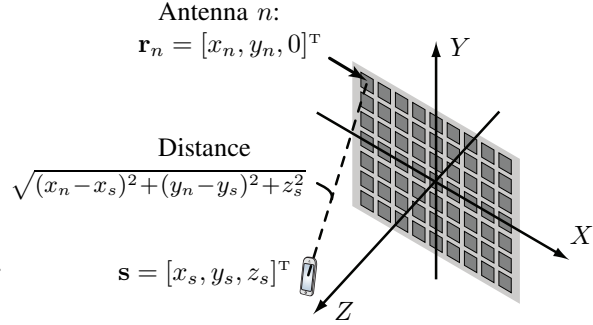}
	\put(-15,10){$\vect{s}= [x_s, y_s, z_s]^{\Ttran}$}
	\put(-5,75){Antenna $n$:}
	\put(-10,67){$\vect{r}_n= [x_n, y_n, 0]^{\Ttran}$}
	\put(-5,41){Distance}
	\put(-42,32){\small $\sqrt{(x_n\!-\!x_s)^2\!+\!(y_n\!-\!y_s)^2\!+\!z_s^2}$}
	\put(98,19){$X$}
	\put(68,66){$Y$}
	\put(39,0){$Z$}
\end{overpic} 
                \caption{A source at an arbitrary location $(x_s,y_s,z_s)$ transmits to the LIS located in the $XY$-plane.} \vspace{-5mm}
                \label{figure_geometricsetup}
\end{figure}
\section{System model}
\label{sec:preliminaries}
 \begin{figure*}[t!]
\begin{align} 
\zeta_{\vect{s},\vect{r}_n} = \frac{1}{4\pi} \sum_{x \in \mathcal{X}_{s,n} } \sum_{ y \in \mathcal{Y}_{s,n} } &\left( \frac{\frac{xy}{z_s^2}}{3 \left(\frac{y^2}{z_s^2}+1\right)\sqrt{ \frac{x^2}{z_s^2}+\frac{y^2}{z_s^2}+1}} 
+ \frac{2}{3} \tan^{-1} \left(  \frac{\frac{xy}{z_s^2}}{\sqrt{ \frac{x^2}{z_s^2}+\frac{y^2}{z_s^2}+1}}
\right) \right) \label{eq:channel-gain-general-case}\tag{3}
\end{align}
\vspace{-.1cm}
\hrule
\vspace{-.2cm}
 \end{figure*}
 
We consider the LIS shown in Fig.~\ref{figure_geometricsetup} consisting of $N$ antennas that each has area $A$. The antennas have size $\sqrt{A} \times \sqrt{A}$ and are equally spaced on a $\sqrt{N} \times \sqrt{N}$ grid. The antennas are deployed edge-to-edge, thus the total area of the LIS is $NA$.  The LIS is centered around the origin in the $XY$-plane. If we number the antennas from left to right, row by row, according to Fig.~\ref{figure_geometricsetup}, the $n$th receive antenna for $n=1,\ldots,N$ is located at $\vect{r}_n= [x_n, y_n, 0]^{\Ttran}$
where 
\vspace{-.0cm}
\begin{align} \label{eq:xn}
x_n &= - \frac{(\sqrt{N}-1)\sqrt{A}}{2} + \sqrt{A} \, \mod(n-1,\sqrt{N}) \\
y_n &= \frac{(\sqrt{N}-1)\sqrt{A}}{2} - \sqrt{A} \left\lfloor \frac{n-1}{\sqrt{N}} \right\rfloor. \label{eq:yn}
\end{align}

\subsection{Channel model}

The following lemma comes from~\cite{bjornson2020power} and extends prior work in \cite{Dardari2019a} to provide a general way of computing channel gains to each of the $N$ antenna elements of the LIS.

\begin{lemma}
\label{lemma1}
Consider a lossless isotropic antenna located at $\vect{s}= [x_s, y_s, z_s]^{\Ttran}$
that transmits a signal that has polarization in the $Y$ direction when traveling in the $Z$ direction. The free-space channel gain $\zeta_{\vect{s},\vect{r}_n}$ at the $n$th receive antenna, located at $\vect{r}_n= [x_n, y_n, 0]^{\Ttran}$, is given by \eqref{eq:channel-gain-general-case} (at top of next page)
where 
\setcounter{equation}{3} 
\begin{align}
\mathcal{X}_{s,n} &= \left\{ \sqrt{A}/2+x_n-x_s,\sqrt{A}/2-x_n+x_s \right\}\end{align} \begin{align} \mathcal{Y}_{s,n} &= \left\{ \sqrt{A}/2+y_n-y_s,\sqrt{A}/2-y_n+y_s \right\}.
\end{align}
\end{lemma}

Lemma~\ref{lemma1} is important when quantifying the channel gain in the so-called geometric near-field of the array \cite{Garcia2019a,Ellingson2019a},\footnote{Note that we assume throughout this paper that $||{\bf s} - {\bf r}_n|| \gg \lambda$, so the system does not operate in the reactive near-field of the transmit antenna (even if it is in the geometric near-field of the array). In fact, this assumption must be made to derive the expression in Lemma~\ref{lemma1}; see \cite{Dardari2019a} for details.} because it takes into account the three fundamental properties that makes it different from the far-field: 1) the distance to the elements varies over the array; 2) the effective antenna areas vary since the element are seen from different angles; 3) the loss from polarization mismatch varies since the signals are received from different angles.
We will use Lemma~\ref{lemma1} in the remainder.

\subsection{Signal model}
\begin{figure}[t!]\hspace{1.5cm}
	\begin{overpic}[width=.7\columnwidth,tics=10]{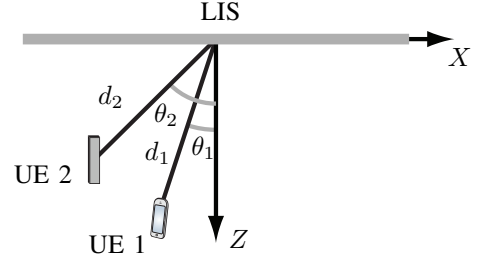}
	\put(42,50){LIS}
	\put(30,20){$d_1$}
	\put(20,32){$d_2$}
	\put(32,28){$\theta_2$}
	\put(40,21){$\theta_1$}
	\put(48,1){$Z$}
	\put(95,40){$X$}
	\put(18,0){UE 1}
	\put(2,15){UE 2}
\end{overpic} 
                \caption{The two UEs are located in the $XZ$-plane at distances $d_k$ for $k=1,2$ and have angles $\theta_k$ for $k=1,2$.}  \vspace{-0.5cm}
                \label{figure_figure_from_above}  
\end{figure}

We consider two single-antenna UEs that communicate with the LIS in Fig.~\ref{figure_geometricsetup} under the following assumption, shown in Fig.~\ref{figure_figure_from_above}. This setup is sufficient to demonstrate a few key results. 

\begin{assumption} \label{assumption3}
UE $k$ for $k=1,2$
is located in the $XZ$-plane at distance $d_k$ from the center of the array with angle $\theta_k \in [-\pi/2,\pi/2]$. Both UEs send a signal that has polarization in the $Y$ direction when traveling in the $Z$ direction.
\end{assumption}

We denote by $\vect{h}_{k} = [h_{k1},\ldots,h_{kN}]^{\Ttran} \in \mathbb{C}^N$ for $k=1,2$ the channel between UE~$k$ and the LIS. Particularly, $h_{kn}=|h_{kn}| e^{-j\phi_{kn}}$ is the channel from the source to the $n$th receive antenna with $|h_{kn}|^2 \in [0,1]$ being the channel gain and $\phi_{kn} \in [0,2\pi)$ the phase shift. Following the geometry stated in Assumption~\ref{assumption3}, the two UEs are located at (see Fig.~\ref{figure_figure_from_above})
\begin{align} 
\vect{s}_1& = [x_{s_{1}}, y_{s_{1}}, z_{s_{1}}]^{\Ttran}=[d_1 \sin(\theta_1), 0, d_1 \cos(\theta_1)]^{\Ttran}\\
\vect{s}_2&= [x_{s_{2}}, y_{s_{2}}, z_{s_{2}}]^{\Ttran}=[d_2 \sin(\theta_2), 0, d_2 \cos(\theta_2)]^{\Ttran}.
\end{align}
From Lemma~\ref{lemma1}, the following corollaries are found.

\begin{corollary}[Exact model]  \label{cor:near-field-model-mMIMO}
Under Assumption~\ref{assumption3}, the channel $h_{kn}=|h_{kn}| e^{-j\phi_{kn}}$ to the $n$th receive antenna is obtained as 
\begin{equation} 
|h_{kn}|^2 = \zeta_{{\bf s}_k,{\bf r}_n}, \hspace{.2cm} \phi_{kn} = 2\pi \, \mod \Bigg( \frac{{||\vect{s}_k- \vect{r}_n||}}{\lambda}  ,1\Bigg).
\end{equation}
\end{corollary}

\begin{corollary}[Far-field approximation]  \label{cor:far-field-model-mMIMO}
Under Assumption~\ref{assumption3}, if UE $k$ is in the geometric far-field of the array, in the sense that $d_k \cos(\theta_k) \gg \sqrt{N A}$, then $h_{kn}=|h_{kn}| e^{-j\phi_{kn}}$ is obtained as
\begin{equation} 
|h_{kn}|^2  \!=\! A\frac{\cos(\theta_{k}) }{4\pi d_{k}^2}, \hspace{.1cm} \phi_{kn} \!=\! 2\pi \, \mod \Big( \frac{d_k\!-\!x_n\sin(\theta_{k})}{\lambda}  ,1\!\Big).
\end{equation}
\end{corollary}


The received signal ${\bf y}\in \mathbb{C}^{M\times 1}$ at the LIS is $\vect{y}=  s_{1}\vect{h}_{1}+ s_{2}\vect{h}_{2} + \vect{n}$
where $s_{i} \sim \CN({0}, p_i)$ is the data signal from UE~$i$ and $\vect{n} \in \mathbb{C}^{M \times N}$ is thermal noise with i.i.d.\ elements distributed as $\CN(0, \sigma^{2})$. We define the average received signal-to-noise ratio (SNR) of UE $i$ as ${\mathsf{ SNR}}_i = {p_i}/{\sigma^{2}}$. The channels are deterministic and thus can be estimated arbitrarily well from pilot signals. Hence, perfect channel state information is assumed. The impact of imperfect knowledge of the interfering UE channel will be investigated in Section~\ref{Sec3.D}.

\begin{figure*}[ht!]
\begin{align} \notag
||\vect{h}_{k}||^{2} = \xi_{d_k,\theta_k,N} 
= & \sum_{i=1}^{2} \Bigg( \frac{ B_k +(-1)^i \sqrt{B_k} \tan(\theta_k)  }{6 \pi (B_k+1)\sqrt{ 2B_k+\tan^2(\theta_k) + 1 + 2(-1)^i \sqrt{B_k} \tan(\theta_k)} } \\ &+ \frac{1}{3\pi} \tan^{-1} \Bigg(  \frac{ 
B_k +(-1)^i \sqrt{B_k} \tan(\theta_k) 
  }{
  \sqrt{2B_k+\tan^2(\theta_k) + 1 + 2(-1)^i \sqrt{B_k} \tan(\theta_k)}
  }
\Bigg) \Bigg) \tag{15}\label{eq:xi-mMIMO}
\end{align}
\hrule
\vspace{-.6cm}
\end{figure*}
 
To detect $s_{1}$ from $\vect{y}$, the LIS uses the combining vector $\vect{v}_{1} \in \mathbb{C}^{M}$, multiplied by the vector $\vect{y}$. By treating the interference as noise, the  SE for UE $1$ is $\mathsf{SE}_1 = \log_2\left(1+\gamma_{1} \right)$
where 
\begin{align}\label{eq:uplink-instant-SINR_case_study}
\gamma_{1} 
= \frac{ \mathsf{SNR}_1|  \vect{v}_{1}^{\Htran} {\vect{h}}_{1}|^2  }{ 
 \mathsf{SNR}_2|  \vect{v}_{1}^{\Htran} {\vect{h}}_{2}|^2 + ||{\vect{v}}_{1}||^{2}
} 
\end{align}
is the signal-to-interference-and-noise ratio (SINR).
We begin by considering MR combining, defined as $\vect{v}_{1} = \vect{h}_{1}/||\vect{h}_{1}||$, leading to   \vspace{-.1cm}
\begin{align}
\gamma_{1}^{\rm{MR}}\!=\!\!
\frac{ \mathsf{SNR}_1 ||\vect{h}_{1}||^{2} }{ 
 \mathsf{SNR}_2\frac{{|  \vect{h}_{1}^{\Htran} {\vect{h}}_{2}|^2}}{||\vect{h}_{1}||^{2}} + 1}\!\!=\! \mathsf{SNR}_1 ||\vect{h}_{1}||^{2}\left(1-\alpha^{\rm{MR}} \right)
\label{eq:uplink-instant-SINR_case_study_1_MR}
\end{align}
with $\alpha^{\rm{MR}} = \frac{\mathsf{SNR}_2\frac{| \vect{h}_{1}^{\Htran} {\vect{h}}_{2}|^2}{||\vect{h}_{1}||^{2}}}{1 +   \mathsf{SNR}_2\frac{{|  \vect{h}_{1}^{\Htran} {\vect{h}}_{2}|^2}}{||\vect{h}_{1}||^{2} } } \vspace{-.1cm}$.
The term
\begin{align} 
\!\!\!\mathsf{SNR}_2\frac{{|  \vect{h}_{1}^{\Htran} {\vect{h}}_{2}|^2}}{||\vect{h}_{1}||^{2}} = \mathsf{SNR}_2 \frac{\left|\sum_{n=1}^N |h_{1n}||h_{2n}|e^{j(\phi_{1n}-\phi_{2n})}\right|^2}{||\vect{h}_{1}||^{2}}
\end{align} 
accounts for the interference generated by UE 2 whereas $\mathsf{SNR}_1  ||\vect{h}_{1}||^{2}$ in~\eqref{eq:uplink-instant-SINR_case_study_1_MR} represents the received SNR in the absence of interference. Since MR does not do anything against the interference, the term $\alpha^{\rm{MR}}$ in \eqref{eq:uplink-instant-SINR_case_study_1_MR} must be interpreted as the performance loss due to the presence of UE 2.

Instead of using the suboptimal MR combining, we note that $\gamma_{1}$ in \eqref{eq:uplink-instant-SINR_case_study} is a generalized Rayleigh quotient with respect to ${\bf v}_1$ and thus is maximized by MMSE combining:  \vspace{-.1cm}
\begin{align}\label{eq:MMSE}
{\bf v}_1 = \Bigg(  \sum\limits_{i=1}^2 \mathsf{SNR}_i{\vect{h}}_{i} {\vect{h}}_{i}^{\Htran} +\vect{I}_{M}  \Bigg)^{\!-1}\!\! {\vect{h}}_{1}
\end{align}
leading to
\begin{align}
\gamma_{1}^{\rm{MMSE}}  = \mathsf{SNR}_1||\vect{h}_{1}||^{2} \big( 1 - \alpha^{\rm{MMSE}} \big)\label{eq:MMSE-max-SINR-2}
\end{align}
with $\alpha^{\rm{MMSE}} = \frac{\mathsf{SNR}_2\frac{| \vect{h}_{1}^{\Htran} {\vect{h}}_{2}|^2}{||\vect{h}_{1}||^{2}}}{1+\mathsf{SNR}_2 ||\vect{h}_{2}||^{2}}$. 

The SINR expression in \eqref{eq:MMSE-max-SINR-2} contains some of the same terms as in \eqref{eq:uplink-instant-SINR_case_study_1_MR}, but has a different structure which makes it behave differently. Unlike $\alpha^{\rm{MR}}$ in~\eqref{eq:uplink-instant-SINR_case_study_1_MR}, the term $\alpha^{\rm{MMSE}}$ in \eqref{eq:MMSE-max-SINR-2} must be interpreted as the performance loss encountered with MMSE when cancelling the interference from UE 2.


\section{Near- and Far-Field Analysis}
\label{sec:power-scaling-laws}
We now analyze the system in the near- and far-field cases. We begin by reviewing the behavior of the channel gain \cite{bjornson2020power}, and then we will look into the interference gain and spectral efficiency. From \cite[Prop.~1]{bjornson2020power}, we have the following result.
\begin{proposition} \label{prop:SNR-mMIMO2}
Under Assumption~\ref{assumption3}, we have that $||\vect{h}_{k}||^{2}  =  \xi_{d_k,\theta_k,N}$
where $\xi_{d_k,\theta_k,N} $ is given by \eqref{eq:xi-mMIMO}
with 
\setcounter{equation}{15} 
\begin{align}\label{eq:SNR_mMIMO_general}
B_k = \frac{NA}{4 d_k^2 \cos^2(\theta_k)}.
\end{align}
In the far-field case, it becomes 
\begin{align} \label{eq:Ch_gain-Far_Field}
	||\vect{h}_{k}||^{2} = N A\frac{\cos(\theta_{k}) }{4\pi d_{k}^2}.
\end{align}  
\end{proposition}
%

\begin{figure} 
        \centering
	\begin{overpic}[width=1.1\columnwidth,tics=10]{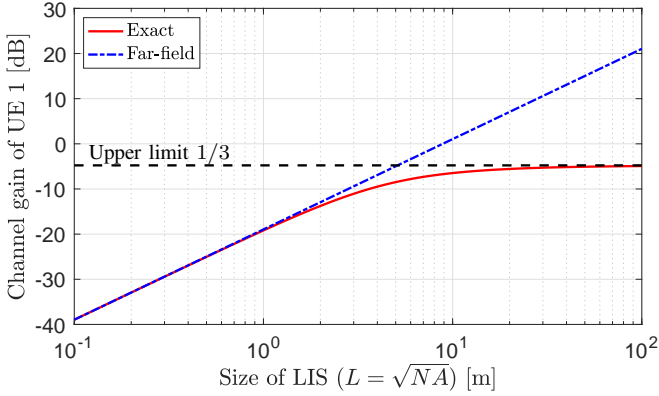}
	\put(15,32){\footnotesize{Upper limit $1/3$}}
\end{overpic} 
        \caption{Behavior of the channel gain $||\vect{h}_{1}||^{2}$ using either the exact model or the far-field approximation. The desired UE 1 has $\theta_1=2^\circ$ and is at a distance $ d_1= 25 \lambda= 2.5$\,m.}\vspace{-0.3cm}
        \label{fig:channel-gain}   
\end{figure}


Note that both \eqref{eq:xi-mMIMO} and \eqref{eq:Ch_gain-Far_Field}  depend on the total LIS area, $NA$. Hence, the channel gain is independent of the wavelength. Since practical elements are sub-wavelength-sized, the number of elements $N$ that is needed to achieve a given channel gain is inversely proportional to $\lambda^2$ .

\begin{figure} 
        \centering
	\begin{overpic}[width=1.1\columnwidth,tics=10]{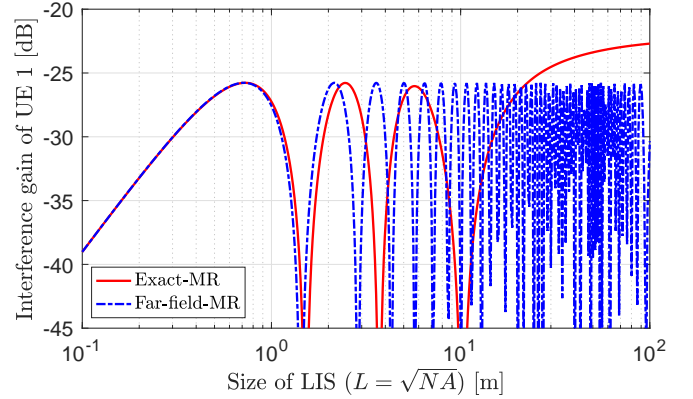}
\end{overpic} 
        \caption{Behavior of the interference gain $\frac{{|  \vect{h}_{1}^{\Htran} {\vect{h}}_{2}|^2}}{||\vect{h}_{1}||^{2} } $ using either the exact model or the far-field approximation. The desired UE 1 is at $\theta_1=2^\circ$ while the interfering UE 2 is at $\theta_2=-2^\circ$. Both UEs are at a distance $ d_1=d_2 = 2.5$\,m. Note that the valleys not always reaching their correct value, $-\infty$ dB, due to the limited numerical precision.}\vspace{-0.4cm}
        \label{fig:interference-analysis} 
\end{figure}

Fig.~\ref{fig:channel-gain} shows the channel gain $||\vect{h}_{1}||^{2}$ as a function of the size of the LIS, i.e., $L = \sqrt{NA}$, using either the exact expression in~\eqref{eq:xi-mMIMO} or the far-field
approximation in~\eqref{eq:Ch_gain-Far_Field}. We consider a setup with $\lambda =0.1$\,m in which UE 1 has $\theta_1 =2^\circ$ and  $d_1= 25 \lambda=2.5$\,m. The results of Fig.~\ref{fig:channel-gain} show that an LIS larger than $L = 1$ m is already enough to notice the approximation gap, whereas  $L \ge 10$ m is needed to approach the upper limit of $1/3$ (this limit comes from considering the polarization mismatch loss~\cite{bjornson2020power} and differs from the $1/2$ limit in~\cite{Hu2018a}). This shows the importance of properly modeling the near-field.

 \vspace{-1mm}
\subsection{Interference gain} \vspace{-1mm}
We now analyze the normalized interference gain $\frac{{|  \vect{h}_{1}^{\Htran} {\vect{h}}_{2}|^2}}{||\vect{h}_{1}||^{2}}$. We notice that the computation of a closed-form expression with the exact channel model is challenging while it takes the simple form with the far-field approximation~\cite[Eq. (12)]{bjornson2019Asilomar}
\begin{equation}
\frac{{|  \vect{h}_{1}^{\Htran} {\vect{h}}_{2}|^2}}{||\vect{h}_{1}||^{2}} = \frac{A\cos(\theta_2)}{4\pi d_2^2}\left|\frac{\sin( \pi L \Omega /\lambda)}{\sin( \pi \sqrt{A} \Omega /\lambda) }\right|^2
\end{equation}
with $L = \sqrt{N A}$ and $\Omega = \sin(\theta_2) - \sin(\theta_1)$. Fig.~\ref{fig:interference-analysis} plots the the normalized interference gain $\frac{{|  \vect{h}_{1}^{\Htran} {\vect{h}}_{2}|^2}}{||\vect{h}_{1}||^{2}}$ as a function of $L$, using either the exact model and the far-field
approximation from Corollaries~\ref{cor:near-field-model-mMIMO} and~\ref{cor:far-field-model-mMIMO}, respectively. We consider a setup with $\lambda =0.1$\,m in which the two UEs have different angles $\theta_1 =2^\circ$ and $\theta_2 = -2^\circ$, but have the same distance from the LIS, given by $d_1=d_2 = 2.5$\,m. In line with the results of Fig.~\ref{fig:channel-gain}, the gap between the exact model and the far-field
approximation is noticeable when $L \ge 1$\,m. We notice that  $L > 100$\,m is needed with the exact model to approach its upper limit. This is more than $3$\,dB higher than the upper limit with the far-field approximation.

\subsection{Spectral efficiency analysis}

\begin{figure} [t!]

        \centering
        \begin{subfigure}[t]{\columnwidth} \centering 
	\begin{overpic}[width=1.1\columnwidth,tics=10]{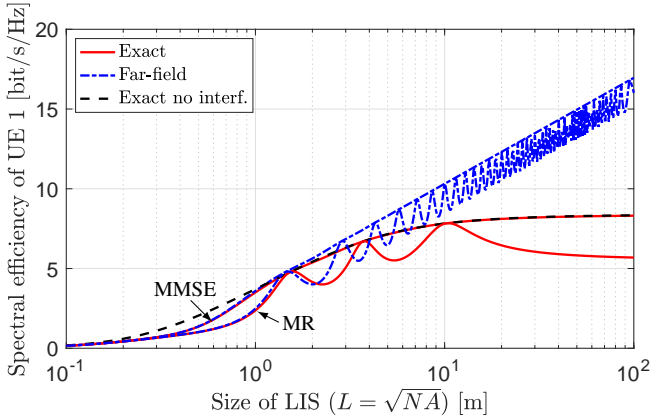}
	\put(30,16){\vector(1, -1){3}}
	\put(25,16.5){\footnotesize{MMSE}}
	\put(42.5,12){\footnotesize{MR}}
	\put(42,13){\vector(-1, .5){3}}
\end{overpic} 
                \caption{Spectral efficiency versus $L = \sqrt{NA}$} \vspace{0mm}
                \label{fig:SE_vs_size} 
        \end{subfigure} 
        \begin{subfigure}[t]{\columnwidth} \centering  
	\begin{overpic}[width=1.1\columnwidth,tics=10]{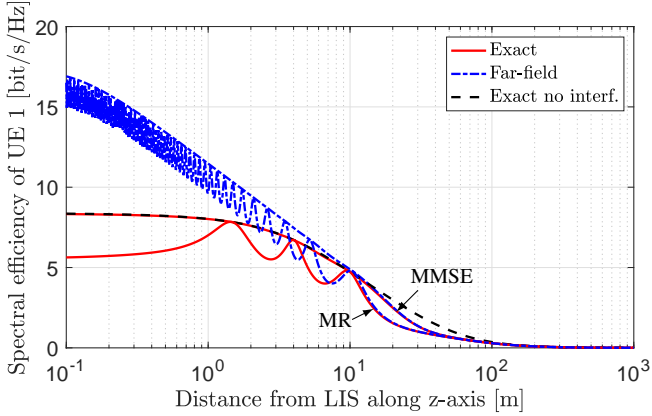}
	\put(61,17){\vector(-1, -1){3}}
	\put(60,18){\footnotesize{MMSE}}
	\put(47.5,12){\footnotesize{MR}}
	\put(52,13){\vector(2, 1){3}}
\end{overpic} 
                \caption{Spectral efficiency versus $z=z_{s_{1}}=z_{s_{2}}$ when $L=6$\,m} 
                \label{fig:SE_vs_dx}
        \end{subfigure} 
        \caption{SE behavior with MR and MMSE. The SE in the ideal interference-free case is reported as reference.}\vspace{-0.6cm}
        \label{fig:SE_analysis}
\end{figure}

Fig.~\ref{fig:SE_vs_size} plots the SE achieved by UE 1 with MR and MMSE as a function of $L$, using either the exact model or the far-field approximation from Corollaries~\ref{cor:near-field-model-mMIMO} and~\ref{cor:far-field-model-mMIMO}. We consider the same setup of Fig.~\ref{fig:interference-analysis}; that is, $\lambda=0.1$\,m, $\theta_1=2^\circ$, $\theta_2=-2^\circ$ and $ d_1=d_2 = 2.5$\,m. We set $p_1=p_2= 30$ dBm and $\sigma^2 = 0$\,dBm. The SE $\log_2 (1 + \mathsf{SNR}_1 ||\vect{h}_{1}||^{2})$ computed with the exact model in the ideal interference-free case is also reported as a reference. Fig.~\ref{fig:SE_vs_size} shows that the SE saturates with both schemes as $L$ increases. However, while MMSE quickly converges to the interference-free case, the performance gap with MR is substantial with the exact model, while it asymptotically vanishes with the far-field model as in \cite[Sec. III-D]{bjornson2019Asilomar}. 

\begin{figure} [t!]
\vspace{-0.3mm}
        \centering
        \begin{subfigure}[t]{\columnwidth} \centering 
	\begin{overpic}[width=0.9\columnwidth,tics=10]{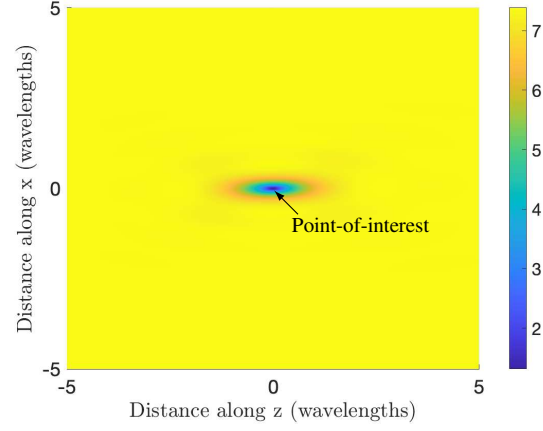}
		\put(50,35){\vector(-1, 1){4}}
	\put(49,32){\footnotesize{Point-of-interest}}
\end{overpic} 
                \caption{SE with MMSE} \vspace{-0.5mm}
                \label{fig:SE_focusing_MMSE} 
        \end{subfigure}
        \begin{subfigure}[t]{\columnwidth} \centering  
	\begin{overpic}[width=0.9\columnwidth,tics=10]{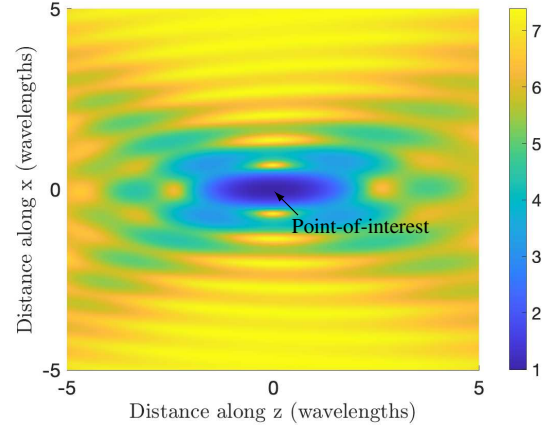}
			\put(50,35){\vector(-1, 1){4}}
	\put(49,32){\footnotesize{Point-of-interest}}
\end{overpic} 
                \caption{SE with MR} 
                \label{fig:SE_focusing_MR}
        \end{subfigure} 
        \caption{SE in bit/s/Hz achieved by UE $1$ when the interfering UE $2$ is transmitting at different locations over the $XZ-$plane. Both MMSE and MR are considered.}\vspace{-0.5cm}
        \label{fig:SE_focusing}
\end{figure}

%

Fig.~\ref{fig:SE_vs_dx} plots the SE achieved by UE 1 with MR and MMSE as a function of  $z=z_{s_{1}}=z_{s_{2}}$, using either the exact model or the far-field approximation. We consider an LIS of size $L=6$\,m, and set $p_1=p_2= 30$\,dBm and $\sigma^2 = 0$\,dBm. MMSE provides the same SE as in the interference-free case for $z \le 10$\,m, whereas it is lower for larger values. The SE gap between MMSE and MR is relatively large for the values of $z$ of practical interest, i.e., smaller than tens of meters. MMSE converges to MR only when $z \ge 40$\,m.

To gain further insights into the large performance gap between MMSE and MR in the near-field, Fig.~\ref{fig:SE_focusing} shows the SE in bit/s/Hz when the desired UE $1$ is fixed at $\theta_1=2^\circ$ $ d_1=2.5$\,m and the interfering UE $2$ is transmitting from different locations over the $XZ-$plane, whose distance from the point-of-interest is measured in wavelengths. We assume $\lambda = 0.1$\,m and use an LIS of size $L=6$\,m. Fig.~\ref{fig:SE_focusing_MMSE} reveals that the SE with MMSE is low only in an elliptic region around the point-of-interest, whose semi-major axis (along $z$ direction) is roughly half-a-wavelength. 
This means that MMSE can efficiently reject any interfering signal that comes from a location that is at least half-a-wavelength away. On the contrary, Fig.~\ref{fig:SE_focusing_MR} shows that a low SE is achieved with MR for most of the locations the interfering UE is transmitting from. This is because MR does not do anything against the interference. The region where the SE achieves its minimum value is still an ellipse but with a larger area. We see that the SE with both MMSE  and MR ranges from $\approx 7$ to $1$ bit/s/Hz (a reduction of $86\%$), but while in the MMSE case the highest values are located in most of the observed area, that is not the case when using MR.

\begin{figure} 
        \centering
	\begin{overpic}[width=1.1\columnwidth,tics=10]{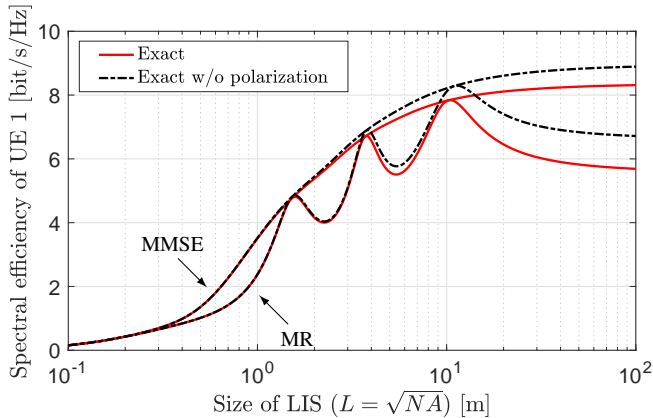}
	\put(28,22){\vector(1, -1){4}}
	\put(23,23){\footnotesize{MMSE}}
	\put(43,13){\vector(-1, 1){4}}
	\put(42,10){\footnotesize{MR}}
\end{overpic} 
        \caption{Impact of polarization on the SE and interference gain of UE $1$ for the same setup in Fig.~\ref{fig:SE_vs_size}.}\vspace{-0.6cm}
        \label{fig:SE_Polarization} 
\end{figure}
\subsection{Impact of polarization mismatch}

Unlike~\cite{Hu2018a}, the analysis and results above take into account the polarization mismatch, which varies the received power of each antenna element since the signals are received from different angles~\cite{bjornson2020power}. This has a double impact on the system performance. Firstly, the global channel gain is reduced and converges to $1/3$ instead of $1/2$ as $L$ grows large; see Fig.~\ref{fig:channel-gain}. Secondly, the mutual interference between UEs changes due to the varying power footprint induced by the polarization mismatch. To quantify the joint effect of these two facts, Fig.~\ref{fig:SE_Polarization} considers the same setup as in Fig.~\ref{fig:SE_vs_size} and compares the SE obtained with the exact model when ignoring or considering the polarization mismatch loss. 
The two effects have a non-negligible impact on both MMSE and MR when $L$ is large. For example, for $L=100$ the SE reduction is $6.5\%$ with MMSE, whereas it is $15\%$ with MR. However, for values of $L$ of practical interest, i.e., in the range between $L=1$ and $L=10$ meters, the polarization effects loss is limited to $5\%$. 


\vspace{-1mm}

\subsection{Imperfect knowledge of interfering UE channel}\label{Sec3.D}
\begin{figure}
\centering
  \includegraphics[width=1.1\linewidth]{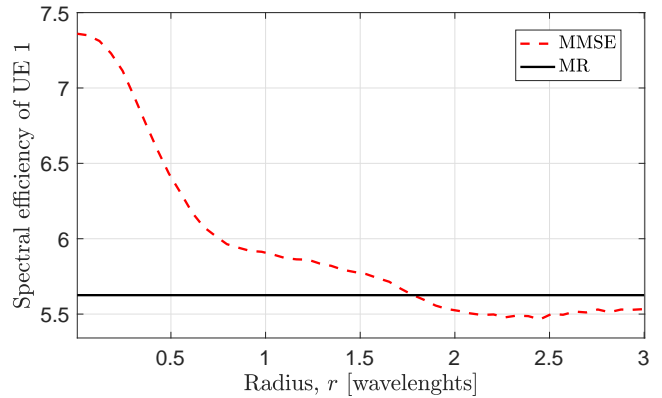}  
  \caption{SE with the exact model when the position of the interfering UE is uncertain within a radius $r$, measured in wavelenghts. We assume $\lambda = 0.1$, $L=6$\,m and an inter-UE distance is $17.5$\,cm$=1.75\,\lambda$.}\vspace{-0.4cm}
  \label{fig:er_ch_interf}	
\end{figure}

The SE analysis above reveals that MMSE performs much better than MR in both the near- and far- fields. However, the analysis relies on perfect knowledge of the channel vectors $\{{\bf h}_1,{\bf h}_2\}$. We now evaluate the robustness of MMSE against imperfect knowledge of ${\bf h}_2$ in \eqref{eq:MMSE}. This is done by assuming that the location of UE $2$ for the computation of ${\bf h}_2$ in \eqref{eq:MMSE} is imperfectly known. Particularly, we assume that the estimated position differs from the true one by an error that is uniformly distributed in a circle of radius $r$, centered at the true position.
Fig.~\ref{fig:er_ch_interf} shows the average SE as a function of $r$, expressed in wavelengths with $\lambda = 0.1$\,m, in the same setup of Fig.~\ref{fig:SE_vs_size} with $L=6$\,m. The average is taken with respect to the randomly generated errors. We see that MR is unaffected by the imperfect knowledge of the interfering UE position since MR does not rely on the knowledge of ${\bf h}_2$. On the other hand, the SE with MMSE decreases very fast. In line with Fig.~\ref{fig:SE_focusing_MMSE}, an estimation error of half-a-wavelength is enough for a $65\%$ reduction. This calls for estimation schemes with centimeter accuracy (having $\lambda=10$\,cm) which is far beyond what we can achieve with state-of-the-art solutions in wireless applications.

\section{Summary}
This paper showed that a realistic assessment of the uplink SE achievable by two single-antenna UEs communicating with an LIS requires the use of an exact near-field channel model, whenever the distance of transmitting UEs is comparable with the LIS size, $L$. It also showed that increasing  $L$ unboundedly does not guarantee the suppression of interference with MR combining, especially when the UEs are closely located in space. MMSE combining is still needed to efficiently suppress the interference. Particularly, the SE with MMSE quickly converges to the interference-free case when the UEs are transmitting from a distance of few meters and values of $L$ of practical interest are considered, i.e., in the range $1 \le L \le 10$ meters. However, the SE with MMSE deteriorates fast in the presence of channel estimation errors. Lastly, we showed that the polarization mismatch should not be ignored since it has a non-negligible impact on SE, especially when $L$ grows. 

\vspace{-.03cm}
\bibliographystyle{IEEEtran}

\bibliography{IEEEabrv,refs}

\end{document}